\documentclass[journal]{IEEEtran}
\usepackage{cite}
\usepackage{graphicx}
\usepackage{float}
\usepackage{subfloat}
\usepackage{stfloats}
\ifCLASSOPTIONcompsoc
\usepackage[caption=false,font=normalsize,labelfont=sf,textfont=sf]{subfig}
\else
\usepackage[caption=false,font=footnotesize]{subfig}
\fi
\usepackage{amsmath}
\usepackage{amsthm}
\usepackage{algorithmic}
\usepackage{array}
\usepackage{tikz}
\usepackage{color}
\usepackage{epsfig,latexsym}
\usepackage{indentfirst}
\usepackage{amssymb}
\usepackage{times}
\usepackage{mathtools}
\usepackage{psfrag}
\usepackage{lastpage}
\usepackage{fancyhdr}
\usepackage{color}
\usepackage{amsthm}
\usepackage{bigints}
\usepackage{algorithm}
\usepackage{algorithmic}
\usepackage{setspace}
\usepackage{booktabs}
\usepackage{amsfonts}
\usepackage{epstopdf}
\usepackage{cleveref}
\usepackage{diagbox}
\usepackage{multirow}
\usepackage{flushend}

\theoremstyle{remark}
\newtheorem{theorem}{\quad Theorem}

\allowdisplaybreaks[4]

\begin{document}
\title{A Novel Time-Based Modulation Scheme in Time-Asynchronous Channels for Molecular Communications}
\author{Qingchao Li
\vspace{-0em}
\thanks{Qingchao Li is with the Key Laboratory of Wireless-Optical Communications, Chinese Academy of Sciences, Department of Electronic Engineering and Information Science, University of Science and Technology of China, Hefei, Anhui, 230027, China (e-mail: liqc@mail.ustc.edu.cn).}}
\maketitle

\begin{abstract}
In this paper, a novel time-based modulation scheme is proposed in the time-asynchronous channel for diffusion-based molecular communication systems with drift. Based on this modulation scheme, we demonstrate that the sample variance of information molecules' arrival time approximately follows a noncentral chi-squared distribution. According to its conditional probability density function (PDF), the asynchronous receiver designs are deduced based on the maximum likelihood (ML) detection, with or without background noise in the channel environment. Since the proposed schemes can be applied to the case of transmitting multiple information molecules, simulation results reveal that the bit error ratio (BER) performance improves with the increase of the number of released information molecules. Furthermore, when the background noise is not negligible, our proposed asynchronous scheme outperforms the asynchronous modulation techniques based on encoding information on the time between two consecutive release of information molecules.
\end{abstract}
\begin{IEEEkeywords}
Molecular communications, diffusion, modulation, asynchronous detection, noncentral chi-squared distribution, background noise.
\end{IEEEkeywords}

\vspace{-0em}
\section{Introduction}
\IEEEPARstart{M}OLECULAR communications are the biological-inspired communication paradigms which utilize chemical signals as information carriers instead of electromagnetic waves in nano-networks \cite{zhai2018anti}, bringing about potential applications in plenty of research areas such as biological engineering \cite{nakano2013molecular}, environment monitor \cite{hirabayashi2011design,nakano2012molecular}, drug delivery systems \cite{chahibi2013molecular,femminella2015molecular,chude2017molecular}, industrial and military applications \cite{loscri2014security}. One of the most promising paradigms is diffusion-based molecular communications (MC) \cite{nakano2011molecular}, in which information molecules are released at the transmitter, propagate in a fluid media and are captured by the receiver. In diffusion-based MC, information can be modulated on the properties of molecules, such as the concentration \cite{kuran2012interference}, the types of molecules \cite{kim2013novel} and the release time of information particles \cite{eckford2009timing,farsad2016capacity}.

In this paper, we focus on timing molecular channels, where the information is conveyed in the release time of molecules. In this modulation schemes, the propagation time of each molecules follows inverse Gaussian distribution (in the channel with drift) or l\'{e}vy distribution (in the channel without drift), and the receiver recovers information based on the arrival time of molecules. In previous work, various detection schemes have been proposed for timing molecular channels. In \cite{srinivas2012molecular}, the conditional probability density function (PDF) of information molecules' arrival time is analytically deduced, and then the maximum likelihood (ML) detector is developed for the recovery of the transmitted information. In \cite{murin2017optimal,murin2017diversity,murin2018exploiting}, considering the high computational complexity of ML detectors, a new low-complexity detection algorithm, based on the first arrival (FA) particle among all transmitted molecules, is proposed. It is shown that the error probability performance of the FA detector is very close to that of the ML detector.

The above detectors are all based on an assumption that the clock at the receiver, which is used to measure the arrival time of information molecules, is fully time-synchronized with the clock at the transmitter, which is used to control the release time of information molecules. However, in molecular communications, many biological activities occur asynchronously with others in the same environment. In actual biological systems, time synchronization can be realized by engineered biological cells. For example, quorum sensing can be utilized to induce synchronization in limit-cycle oscillators \cite{schwab2012dynamical}, genetic relaxation oscillators \cite{taylor2009dynamical} and synthetic gene oscillators \cite{mcmillen2002synchronizing}. Hormones and signaling protein molecules can be employed to synchronize activities between different nano-systems in a molecular communication network \cite{chude2017molecular}.

When it comes to deal with the issue of time synchronization from the aspect of modulation/demodulation schemes, previous work mainly focus on the following three points.

Firstly, time synchronization in MC systems can be realized by adding training sequences. In \cite{lin2015diffusion}, a two-way message exchange mechanism is proposed, where the clock offset between the transmitter and the receiver is measured by maximum likelihood estimation. In \cite{lin2016time}, a one-way time synchronization
scheme is proposed to estimate the clock offset between the nanomachines of the transmitter and the receiver by Newton-Raphson method. However, adding training sequences to estimate the clock offset requires additional energy consumption of the MC system.

Secondly, two types of information molecules utilized in modulation schemes provides a feasible way for handling the issue of time synchronization in MC systems. In \cite{haselmayr2019impact}, the binary molecule shift keying modulation (MoSK) scheme is employed, which maps bit-0 and bit-1 to a single molecule of type-$a$ and type-$b$ respectively. The receiver conducts an asynchronous demodulation scheme which exploits the arrival order of information molecules instead of the arrival time. The MoSK scheme is also employed in \cite{lin2015asynchronous}, while in which bit-0 and bit-1 are mapped to multiple molecules of type-$a$ or type-$b$ respectively. And information is recovered at the receiver based on an asynchronous detection algorithm which simply counts and compares the number of each type of molecules received. Furthermore, since each bit is represented by multiple molecules, this asynchronous scheme can operate in the channel environment with background noise. In \cite{atakan2012nanoscale}, Molecular ARray-based COmmunication (MARCO) scheme is proposed, where bit-0 is conveyed by emitting information particle $a$ followed by information particle $b$, and bit-1 is conveyed by emitting information particle $b$ followed by information particle $a$. At the receiver, the arrival order of these two particles is used to recover the information conveyed without any need for time synchronization. However, the above schemes employe two types of information molecules, which increases the complexity of the receiver because it requires the reception to have the ability to identify the type of received molecules.

Lastly, there is an asynchronous modulation technique which employs only one type of molecules and does not use training messages for estimating clock offset. The work \cite{farsad2016impact,farsad2017communication} propose an asynchronous modulation scheme, in which information bit is conveyed based on the time interval between two released molecules. Furthermore, the conditional PDF of the arrival time interval of two molecules is deduced, based on which the maximum likelihood optimal decision can be implemented asynchronously. In this scheme, however, since only two molecules per bit are used for modulation, the detection algorithm works only if there is no background noise in the channel environment (i.e., all released information molecules can be absorbed by the receiver without degrading in the channel), which severely restricts the scope of applications in actual communication systems.

In this paper, a novel time-based modulation scheme is proposed by modifying the conventional pulse position modulation (PPM) \cite{garralda2011diffusion}. According to the sample variance of information molecules' arrival time, the receiver recovers the information based on the maximum likelihood criterion. The major contributions of this paper include: 1) A novel modulation scheme and the corresponding asynchronous detection algorithm are proposed without need of time synchronization or clock offset estimation; 2) Only one type of information molecules is employed in the MC system, which simplifies the complexity of the reception; 3) The proposed asynchronous scheme can work in the channel environment with background noise.

The remainder of this paper is organized as follows. Section II presents the physical model and the asynchronous channel model. In Section III, a novel time-based binary modulation scheme is proposed. And the corresponding asynchronous detection algorithm is given in Section IV. The bit error ratio (BER) performance of the proposed scheme is theoretically analyzed in Section V. Numerical results are presented in Section VI. In Section VII, the proposed asynchronous scheme is extend to the case of multi-ary. Finally, the conclusion is drawn in Section VIII.

\vspace{-0em}
\section{System Model}
\vspace{-0em}

\subsection{Physical Model}
\begin{figure}[!t]
\vspace{-0em}
\setlength{\abovecaptionskip}{-0pt}
\setlength{\belowcaptionskip}{-0pt}
    \centering
    \includegraphics[width=3.2in]{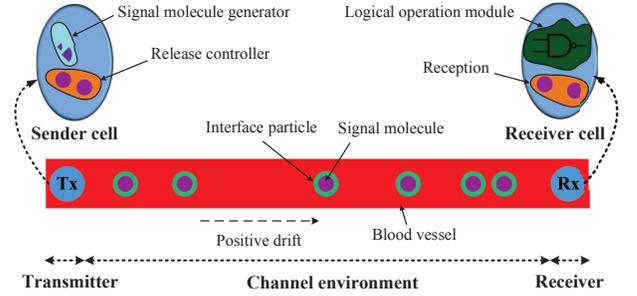}
    \caption{The diagram of the cell-to-cell diffusion-based MC system in blood vessels.}\label{physical_model_fig}
\vspace{-0em}
\end{figure}

In this paper, we assume a cell-to-cell diffusion-based MC system in blood vessels. Fig. \ref{physical_model_fig} illustrates a diagram of the employed physical model, including a transmitter (Tx), channel environment and a receiver (Rx).

\subsubsection{Transmitter}
The transmitter is a sender cell engineered by genetically modifying biological cells. The sender cell includes an information molecule generator \cite{you2004programmed,chen2005artificial,basu2005synthetic}, and a release controller with oscillators \cite{cheong2010oscillatory}.

\begin{itemize}
  \item The information molecule generator in the sender cell can synthesize information molecules such as hexoses \cite{kim2014symbol}.
  \item The release controller emits information molecules into channel environment. Since information is conveyed in the release time of molecules, oscillators are utilized to control the transmitting time of molecules in the release controller.
\end{itemize}

\subsubsection{Channel environment}
The information molecules propagate in the blood vessel, in which the medium has a positive drift velocity. In the channel environment, interface particles (e.g., nano-scale capsules \cite{langer2001drugs}) are employed to protect information molecules from degrading by background noise.

\subsubsection{Receiver}
The receiver is composed of a receiver cell which is synthesized by a biological cell with genetic engineering. The sender cell includes a reception with oscillators and a logical operation module.

\begin{itemize}
  \item The reception in the receiver cell can absorb information molecules. Furthermore, oscillators are equipped in the receiver cell to record the arrival time of each information molecule. Since this is a time-asynchronous scheme, the clock in the sender cell does not need to be synchronized with the clock in the receiver cell, which can significantly reduce system complexity.
  \item The logical operation module, which can be embedded on bio-nanomachines \cite{nakano2013transmission,nakano2014molecular,felicetti2014tcp}, is used for information decoding. In \cite{weiss2003genetic}, logical NAND gates can be designed by mRNA in biological cells, which provides a basic building block for complexity logical operations because NAND gate has the property of functional completeness, i.e., any other logic function (e.g. NOT gate, AND gate, OR gate, etc.) can be implemented by the combination of NAND gates \cite{mano2001logic}.
\end{itemize}

\vspace{-0em}
\subsection{Channel Model}
\begin{figure}[!t]
\vspace{-0em}
\setlength{\abovecaptionskip}{-0pt}
\setlength{\belowcaptionskip}{-0pt}
    \centering
    \includegraphics[width=2.5in]{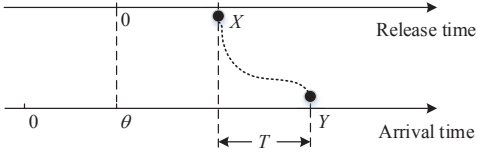}
    \caption{The diagram of time-asynchronous channel model.}\label{channel_model_fig}
\vspace{-0em}
\end{figure}

In this paper, it is assumed that the transmitter and the receiver have their own clocks, between which there is an unknown clock offset, denoted as $\theta$. Fig. \ref{channel_model_fig} illustrates this time-asynchronous channel model, where $X$ is the release time of the information molecule, $T$ is the propagation time from the transmitter to the receiver and $Y$ is the arrival time. Thus, this process can be modeled as
\begin{align}\label{process_model}
    Y=X+T+\theta.
\end{align}

Since the medium has positive drift in the direction of the information propagation and the drift is the major mean of transport, the blood vessel can be modeled as a 1-Dimension channel \cite{tavakkoli2016performance}. The propagation time $T$ is the \textit{first hitting time}, following the inverse Gaussian (IG) distribution \cite{eckford2007nanoscale}:
\begin{align}\label{IG_distribution}
    T\thicksim IG\left(\mu,\lambda\right)=IG\left(\frac{d}{v},\frac{d^2}{2D}\right),
\end{align}
where $\mu$ is the mean and $\lambda$ is the shape parameter of IG distribution, $d$ is the distance between the transmitter and the receiver, $v$ is the drift velocity and $D$ is the diffusion coefficient. In IG distribution, the skewness, a measure of asymmetry, is given by
\begin{align}\label{IG_skewness}
    \gamma=3\left(\frac{\mu}{\lambda}\right)^{1/2}=3\left(\frac{2D}{vd}\right)^{1/2}.
\end{align}
If $vd\gg D$, then $\gamma\approx 0$, which means the propagation time $T$ approximately follows the normal distribution \cite{chhikara1988inverse}:
\begin{align}\label{normal_distribution}
    T\thicksim \mathcal{N}\left(\mu,\sigma^2\right)=\mathcal{N}\left(\mu,\frac{\mu^3}{\lambda}\right)=\mathcal{N}\left(\frac{d}{v},\frac{2Dd}{v^3}\right),
\end{align}
where $\sigma^2$ is the variance of the normal distribution.

\vspace{-0em}
\section{Binary Modulation Scheme}
In this section, a novel binary modulation scheme is proposed for time-asynchronous channel models, and the corresponding multi-ary modulation scheme will be derived in Section VII.

\begin{figure}[!t]
\vspace{-0em}
\setlength{\abovecaptionskip}{-0pt}
\setlength{\belowcaptionskip}{-0pt}
    \centering
    \includegraphics[width=2.5in]{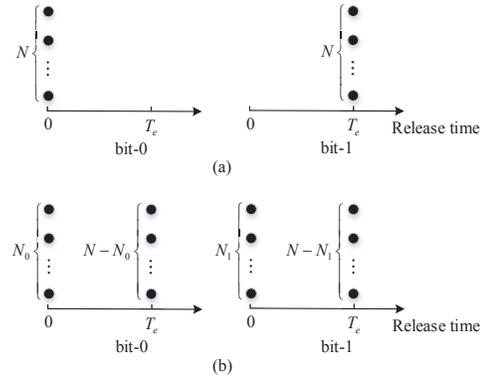}
    \caption{The diagram of (a) the conventional binary PPM scheme and (b) the proposed time-based binary modulation scheme.}\label{modulation_2ary_fig}
\vspace{-0em}
\end{figure}

In the conventional PPM scheme \cite{garralda2011diffusion} illustrated in Fig. \ref{modulation_2ary_fig} (a), information is encoded in the release time of molecules, in which $N$ molecules are simultaneously released at time $X=0$ to convey information bit-0 or at time $X=T_e$ to convey information bit-1. When the PPM scheme is employed in a time-synchronous channel model, the receiver can recover the information effectively according to the arrival time of information molecules. However, in the a time-asynchronous channel model defined in (\ref{process_model}), the conditional joint probability density function of the arrival time of $N$ molecules is given by
\begin{align}\label{PPM_joint_conditional}
    \notag &f_{Y_1,\cdots,Y_N|X}\left(y_1,\cdots,y_N|x\right)\\
    =&\left\{
        \begin{aligned}
            &f_{T_1,\cdots,T_N}\left(t_1-\theta,\cdots,t_N-\theta\right)\qquad\qquad\quad,\mathrm{if}\ X=0\\
            &f_{T_1,\cdots,T_N}\left(t_1-T_e-\theta,\cdots,t_N-T_e-\theta\right)\ ,\mathrm{if}\ X=T_e,
        \end{aligned}
    \right.
\end{align}
in which the conditional joint probability density function of $Y_1,\cdots,Y_N$ given $X=0$ is a time shift of that given $X=T_e$. Since the clock offset $\theta$ is unknown, the receiver can not recover the information in the conventional PPM scheme.

The proposed time-based modulation scheme, which is illustrated in Fig. \ref{modulation_2ary_fig} (b), is an improvement of the conventional PPM scheme. When information bit-0 is conveyed, $N_0$ molecules are released at time $X=0$ and $N-N_0$ molecules are released at time $X=T_e$; when information bit-1 is conveyed, $N_1$ molecules are released at time $X=0$ and $N-N_1$ molecules are released at time $X=T_e$. The best selection of $N_0$ and $N_1$ (i.e., with the minimum error probability) is determined by the value of $N$, $T_e$ and channel parameters, which can be derived numerically.

\vspace{-0em}
\section{Receiver Designs}
In this section, the receiver designs based on the sample variance of information molecules' arrival time for the proposed time-based binary modulation scheme are deduced in the time-asynchronous channel model.

\vspace{-0em}
\subsection{Receiver Design when Background Noise is Negligible}
When $N$ information molecules are released at the transmitter and propagate in the channel environment where the background noise is negligible, the receiver can capture them without loss. If the information conveyed is bit-0, the release time of the $i$-th molecule, denoted as $X_i$, is given by
\begin{align}\label{X_i_bit0}
    X_i=&\left\{
            \begin{aligned}
                &0\ \ ,i\in\left\{1,2,\cdots,N_0\right\}\\
                &T_e\ ,i\in\left\{N_0+1,N_0+2,\cdots,N\right\};
            \end{aligned}
        \right.
\end{align}
if the information conveyed is bit-1, $X_i$ is given by
\begin{align}\label{X_i_bit1}
    X_i=&\left\{
            \begin{aligned}
                &0\ \ ,i\in\left\{1,2,\cdots,{N_1}\right\}\\
                &T_e\ ,i\in\left\{{N_1}+1,{N_1}+2,\cdots,N\right\}.
            \end{aligned}
        \right.
\end{align}
Thus, the arrival time of each information molecule, denoted as $Y_1,Y_2,\cdots,Y_N$, is given by
\begin{align}\label{Y_i_distribution}
    Y_i=X_i+T_i+\theta,\ i\in\left\{1,2,\cdots,N\right\},
\end{align}
where $T_1,T_2,\cdots,T_N$ is the propagation time of information molecules, each of which has an independent and identically distributed (i.i.d.) normal distribution defined in (\ref{normal_distribution}), that is
\begin{align}\label{T_i_distribution}
    T_i\thicksim \mathcal{N}\left(\mu,\sigma^2\right),\ \forall i\in\left\{1,2,\cdots,N\right\}.
\end{align}

According to the statistical theory \cite{degroot2012probability}, the sample variance of $Y_1,Y_2,\cdots,Y_N$ is defined as
\begin{align}\label{Y_i_variance_definition}
    S^2=\frac{1}{N-1}\sum_{i=1}^{N}\left(Y_i-\overline{Y}_{\left(1:N\right)}\right)^2,
\end{align}
where $\overline{Y}_{\left(1:N\right)}$ is the mean of $Y_1,Y_2,\cdots,Y_N$.

\begin{theorem}\label{theorem_variance_bit0}
When the information conveyed is bit-0, denoted as $B=0$, the sampling distribution of $S^2$ defined in (\ref{Y_i_variance_definition}) is given by
\begin{align}\label{variance_bit0}
    \frac{\left(N-1\right)S^2}{\sigma^2}\thicksim\chi{'}_{N-1}^{2}\left(\lambda_{N_0}^{(N)}\right),
\end{align}
where
\begin{align}\label{sigma_bit0}
    \lambda_{N_0}^{(N)}=\frac{N_0\left(N-N_0\right){T_e}^2}{N\sigma^2},
\end{align}
and $\chi{'}_{N-1}^2(\lambda_{N_0}^{(N)})$ is the noncentral chi-squared distribution with $N-1$ degrees of freedom and non-centrality parameter $\lambda_{N_0}^{(N)}$.
\end{theorem}
\begin{IEEEproof}
When (\ref{Y_i_distribution}) is plugged into (\ref{Y_i_variance_definition}), we can get that
\begin{align}\label{proof_theorem_1_1}
    \notag S^2&=\frac{1}{N-1}\sum_{i=1}^{N}\Big[X_i+T_i+\theta-\frac{1}{N}\sum_{i=1}^{N}\left(X_i+T_i+\theta\right)\Big]^2\\
    \notag&=\frac{1}{N-1}\sum_{i=1}^{N}\Big[X_i+T_i-\frac{1}{N}\sum_{i=1}^{N}X_i-\frac{1}{N}\sum_{i=1}^{N}T_i\Big]^2\\
    \notag&=\frac{1}{N-1}\sum_{i=1}^{N}\Big[X_i+T_i-\overline{X}-\overline{T}\Big]^2\\
    &=\frac{1}{N-1}\sum_{i=1}^{N}\Big[\left(T_i-\overline{T}\right)^2+2\left(X_i-\overline{X}\right)T_i+\left(X_i-\overline{X}\right)^2\Big],
\end{align}
where $\overline{X}$ and $\overline{T}$ are the means of $X_1,X_2,\cdots,X_N$ and $T_1,T_2,\cdots,T_N$ respectively. Thus,
\begin{align}\label{proof_theorem_1_3}
    \notag&\frac{\left(N-1\right)S^2}{\sigma^2}=\sum_{i=1}^{N}\frac{\left(T_i-\overline{T}\right)^2}{\sigma^2}+\sum_{i=1}^{N}\frac{2\left(X_i-\overline{X}\right)}{\sigma^2}T_i\\
    &\qquad\qquad\qquad\quad+\sum_{i=1}^{N}\frac{\left(X_i-\overline{X}\right)^2}{\sigma^2},
\end{align}
wherein
\begin{align}
    \notag&\sum_{i=1}^{N}\frac{\left(X_i-\overline{X}\right)^2}{\sigma^2}\\
    \notag=&\frac{1}{\sigma^2}\left(\sum_{i=1}^{N}{X_i}^2-N\overline{X}^2\right)\\
    \notag=&\frac{1}{\sigma^2}\left(\left(N-N_0\right){T_e}^2-N\left({\frac{\left(N-N_0\right)T_e}{N}}\right)^2\right)\\
    \notag=&\frac{N_0\left(N-N_0\right){T_e}^2}{N\sigma^2}\\
    =&\lambda_{N_0}^{(N)}.
\end{align}

Since $T_i$ ($i=1,2,\cdots,N$) are independent and follows $\mathcal{N}\left(\mu,\sigma^2\right)$, we can get that
\begin{align}\label{proof_theorem_1_5}
   \notag&\sum_{i=1}^{N}\frac{2\left(X_i-\overline{X}\right)}{\sigma^2}T_i+\sum_{i=1}^{N}\frac{\left(X_i-\overline{X}\right)^2}{\sigma^2}\\
   \notag\thicksim&\mathcal{N}\Bigg(\mu\sum_{i=1}^{N}\frac{2\left(X_i-\overline{X}\right)}{\sigma^2}+\sum_{i=1}^{N}\frac{\left(X_i-\overline{X}\right)^2}{\sigma^2},\\
   \notag&\qquad\qquad\qquad\sigma^2\sum_{i=1}^{N}\left[\frac{2\left(X_i-\overline{X}\right)}{\sigma^2}\right]^2\Bigg)\\
   \notag=&\mathcal{N}\Bigg(\sum_{i=1}^{N}\frac{\left(X_i-\overline{X}\right)^2}{\sigma^2},4\sum_{i=1}^{N}\frac{\left(X_i-\overline{X}\right)^2}{\sigma^2}\Bigg)\\
   =&\mathcal{N}\left(\lambda_{N_0}^{(N)},4\lambda_{N_0}^{(N)}\right),
\end{align}
and
\begin{align}\label{proof_theorem_1_4}
    \sum_{i=1}^{N}\frac{\left(T_i-\overline{T}\right)^2}{\sigma^2}\thicksim\chi_{N-1}^2,
\end{align}
where $\chi_{N-1}^2$ is the chi-squared distribution with $N-1$ degrees of freedom.

Let $V_1,V_2,\cdots,V_N$ are independent standard normal random variables. According to (\ref{proof_theorem_1_5}), it can be written that
\begin{align}\label{proof_theorem_7_4}
    \sum_{i=1}^{N}\frac{2\left(X_i-\overline{X}\right)}{\sigma^2}T_i+\sum_{i=1}^{N}\frac{\left(X_i-\overline{X}\right)^2}{\sigma^2}=2\sqrt{\lambda_{N_0}^{(N)}}V_N+\lambda_{N_0}^{(N)}.
\end{align}
According to (\ref{proof_theorem_1_4}), it can be written that
\begin{align}\label{proof_theorem_7_3}
    \sum_{i=1}^{N}\frac{\left(T_i-\overline{T}\right)^2}{\sigma^2}=\sum_{i=1}^{N-1}{V_i}^2.
\end{align}

According to (\ref{proof_theorem_1_3}), (\ref{proof_theorem_7_3}) and (\ref{proof_theorem_7_4}), we can get that
\begin{align}\label{proof_theorem_1_7_5}
    \notag\frac{\left(N-1\right)S^2}{\sigma^2}&=\sum_{i=1}^{N-1}{V_i}^2+2\sqrt{\lambda_{N_0}^{(N)}}V_N+\lambda_{N_0}^{(N)}\\
    \notag&=\sum_{i=1}^{N-1}{V_i}^2-{V_N}^2+{V_N}^2+2\sqrt{\lambda_{N_0}^{(N)}}V_N+\lambda_{N_0}^{(N)}\\
    &=\sum_{i=1}^{N-1}{V_i}^2-{V_N}^2+\left({V_N}+\sqrt{\lambda_{N_0}^{(N)}}\right)^2.
\end{align}
In (\ref{proof_theorem_1_7_5}), according to the property of noncentral chi-squared distribution, it can be gotten that
\begin{align}\label{proof_theorem_1_7_6}
    \sum_{i=1}^{N-1}{V_i}^2-{V_N}^2\thicksim\chi_{N-2}^2=\chi{'}_{N-2}^2\left(0\right),
\end{align}
\begin{align}\label{proof_theorem_1_7_7}
    \left({V_N}+\sqrt{\lambda_{N_0}^{(N)}}\right)^2\thicksim\chi{'}_{1}^2\left(\lambda_{N_0}^{(N)}\right).
\end{align}
According to (\ref{proof_theorem_1_7_5}), (\ref{proof_theorem_1_7_6}) and (\ref{proof_theorem_1_7_7}), the sampling distribution of $\frac{(N-1)S^2}{\sigma^2}$ is given by
\begin{align}\label{proof_theorem_1_8}
    \frac{\left(N-1\right)S^2}{\sigma^2}\thicksim\chi{'}_{N-1}^2\left(\lambda_{N_0}^{(N)}\right).
\end{align}

Thus, when the information conveyed is bit-0, $\frac{\left(N-1\right)S^2}{\sigma^2}$ follows the noncentral chi-squared distribution with $N-1$ degrees of freedom and non-centrality parameter $\lambda_{N_0}^{(N)}=\frac{N_0\left(N-N_0\right){T_e}^2}{N\sigma^2}$. \textit{Theorem} \ref{theorem_variance_bit0} is proved.
\end{IEEEproof}

According to the probability density function of noncentral chi-squared distribution in \cite{abramowitz1965handbook}, the conditional PDF of $\frac{\left(N-1\right)S^2}{\sigma^2}$, given $B=0$, is
\begin{align}\label{S2_bit0_pdf}
    \notag&f_{\frac{\left(N-1\right)S^2}{\sigma^2}|B}\left(z|0\right)\\
    =&\frac{1}{2}e^{-{\frac{z+\lambda_{N_0}^{(N)}}{2}}}\left(\frac{z}{\lambda_{N_0}^{(N)}}\right)^{\frac{N-3}{4}}I_{\frac{N-3}{2}}\left(\sqrt{\lambda_{N_0}^{(N)}z}\right),
\end{align}
where $I_\nu(y)$ is a modified Bessel function of the first kind given by
\begin{align}\label{I_v_y}
    I_\nu(y)=\left(\frac{y}{2}\right)^\nu\sum_{j=0}^{\infty}\frac{\left(y^2/4\right)^j}{j!\Gamma\left(\nu+j+1\right)}.
\end{align}

Similarly, the following theorem can be gotten.
\begin{theorem}\label{theorem_variance_bit1}
When the information conveyed is bit-1, denoted as $B=1$, the sampling distribution of $S^2$ defined in (\ref{Y_i_variance_definition}) is given by
\begin{align}\label{variance_bit1}
    \frac{\left(N-1\right)S^2}{\sigma^2}\thicksim\chi{'}_{N-1}^{2}\left(\lambda_{N_1}^{(N)}\right),
\end{align}
where
\begin{align}\label{sigma_bit1}
    \lambda_{N_1}^{(N)}=\frac{N_1\left(N-N_1\right){T_e}^2}{N\sigma^2}.
\end{align}
\end{theorem}

The conditional PDF of $\frac{\left(N-1\right)S^2}{\sigma^2}$, given $B=1$, is
\begin{align}\label{S2_bit1_pdf}
    \notag&f_{\frac{\left(N-1\right)S^2}{\sigma^2}|B}\left(z|1\right)\\
    =&\frac{1}{2}e^{-{\frac{z+\lambda_{N_1}^{(N)}}{2}}}\left(\frac{z}{\lambda_{N_1}^{(N)}}\right)^{\frac{N-3}{4}}I_{\frac{N-3}{2}}\left(\sqrt{\lambda_{N_1}^{(N)}z}\right).
\end{align}

The receiver observes $N$ information molecules' arrival time $Y_1,Y_2,\cdots,Y_N$, calculates their sample variance $S^2$ according to (\ref{Y_i_variance_definition}), and needs to give $\hat{B}$, an estimation of the conveyed information bit $B$. Given an observation $\frac{\left(N-1\right)S^2}{\sigma^2}=z$, the maximum likelihood estimation of $B$, denoted as $\hat{B}_{ML}$, is given by
\begin{align}\label{ML_receiver_design}
    \hat{B}_{ML}=\arg\max_{b\in\{0,1\}}f_{\frac{\left(N-1\right)S^2}{\sigma^2}|B}\left(z|b\right).
\end{align}

\begin{figure}[!t]
\vspace{-0em}
\setlength{\abovecaptionskip}{-0pt}
\setlength{\belowcaptionskip}{-0pt}
    \centering
    \includegraphics[width=4.5in]{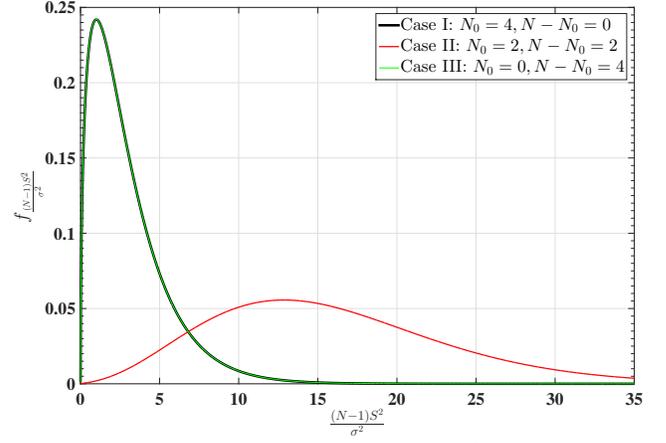}
    \caption{An example of the conditional PDF of $\frac{\left(N-1\right)S^2}{\sigma^2}$ in capillaries, where $N=4$, $T_e=0.1\mathrm{s}$.}\label{variance_illustration}
\vspace{-0em}
\end{figure}

Fig. \ref{variance_illustration} gives an example of the conditional PDF of $\frac{\left(N-1\right)S^2}{\sigma^2}$, where $N=4$, $T_e=0.1\mathrm{s}$, the channel environment is in capillaries and the channel parameters are set according to Table \ref{parameters} in Section VI. In case I, the number of molecules released at time $X=0$ is 4 and released at time $X=T_e$ is 0; in case II, the number of molecules released at time $X=0$ is 2 and released at time $X=T_e$ is 2; in case III, the number of molecules released at time $X=0$ is 0 and released at time $X=T_e$ is 4. In the conventional PPM, essentially, information bit-0 and bit-1 are modulated based on case I and case III. It can be found that the sample variance of information molecules' arrival time in these two cases follows the same sampling distribution, which results in the failure of demodulation in the time-asynchronous channel model. In the proposed modulation scheme, however, information bit-0 and bit-1 can be modulated based on case I and case II (or case II and case III). It can be found that the sample variance of information molecules' arrival time in these two cases follows different sampling distributions, which enables the information demodulated based on the sample variance of molecules' arrival time by maximum likelihood criterion in the time-asynchronous channel model.

\vspace{-0em}
\subsection{Receiver Design when Background Noise is not Negligible}
When the $N$ information molecules propagate in the channel environment where the background noise is not negligible, some of them would degrade and the number of molecules absorbed at the receiver would be less than $N$. We assume that each molecule vanishes in the channel with an equal probability $P_d$ independently.

At the receiver, we denote the number of information molecules absorbed by the receiver as $M$ ($M\leq N$) and their arrival time as $Y_1,Y_2,\cdots,Y_M$ respectively. The simple variance of $Y_1,Y_2,\cdots,Y_M$ is defined as
\begin{align}\label{Y_i_variance_definition_M}
    S^2=\frac{1}{M-1}\sum_{i=1}^{M}\left(Y_i-\overline{Y}_{\left(1:M\right)}\right)^2,
\end{align}
where $\overline{Y}_{\left(1:M\right)}$ is the mean of $Y_1,Y_2,\cdots,Y_M$. Among these $M$ molecules, the number of molecules released at time $X=0$ is a random variable, denoted as $K$, then the number of molecules released at time $X=T_e$ is $M-K$. The random variable $K$ follows hypergeometric distribution: when the information conveyed is bit-0, the probability mass function (pmf) of $K$ is given by
\begin{align}\label{hypergeometric_0}
    p_{K|B}\left(k|0\right)=\frac{\binom{N_0}{k}\binom{N-N_0}{M-k}}{\binom{N}{M}},
\end{align}
where the support of random variable $K$, denoted as $\mathcal{K}_0$, is
\begin{align}
    \mathcal{K}_0:\ K\in\left\{\max\left(0,M+N_0-N\right),\cdots,\min\left(N_0,M\right)\right\};
\end{align}
when the information conveyed is bit-1, the pmf of $K$ is given by
\begin{align}\label{hypergeometric_1}
    p_{K|B}\left(k|1\right)=\frac{\binom{N_1}{k}\binom{N-N_1}{M-k}}{\binom{N}{M}},
\end{align}
where the support of random variable $K$, denoted as $\mathcal{K}_1$, is
\begin{align}
    \mathcal{K}_1:\ K\in\left\{\max\left(0,M+N_1-N\right),\cdots,\min\left(N_1,M\right)\right\}.
\end{align}

Based on (\ref{S2_bit0_pdf}) and (\ref{S2_bit1_pdf}), it can be analogously deduced that when $M$ molecules are received, $k$ of which are released at time $X=0$ and $M-k$ of which are released at time $X=T_e$, the PDF of $\frac{\left(M-1\right)S^2}{\sigma^2}$, denoted as $f_{\frac{\left(M-1\right)S^2}{\sigma^2}}\left(z;k\right)$, is
\begin{align}\label{S2_k_pdf}
    \notag&f_{\frac{\left(M-1\right)S^2}{\sigma^2}}\left(z;k\right)\\
    =&\frac{1}{2}e^{-{\frac{z+\lambda_{k}^{(M)}}{2}}}\left(\frac{z}{\lambda_{k}^{(M)}}\right)^{\frac{M-3}{4}}I_{\frac{M-3}{2}}\left(\sqrt{\lambda_{k}^{(M)}z}\right),
\end{align}
where
\begin{align}\label{sigma_k}
    \lambda_{k}^{(M)}=\frac{k\left(M-k\right){T_e}^2}{M\sigma^2}.
\end{align}

According to (\ref{hypergeometric_0}), (\ref{hypergeometric_1}) and (\ref{S2_k_pdf}), the conditional PDF of $\frac{\left(M-1\right)S^2}{\sigma^2}$, given $B=0$, is
\begin{align}\label{S2_k_bit0_pdf}
    f_{\frac{\left(M-1\right)S^2}{\sigma^2}|B}\left(z|0\right)=\sum_{k\in\mathcal{K}_0}p_{K|B}\left(k|0\right)f_{\frac{\left(M-1\right)S^2}{\sigma^2}}\left(z;k\right),
\end{align}
and the conditional PDF of $\frac{\left(M-1\right)S^2}{\sigma^2}$, given $B=1$, is
\begin{align}\label{S2_k_bit1_pdf}
    f_{\frac{\left(M-1\right)S^2}{\sigma^2}|B}\left(z|1\right)=\sum_{k\in\mathcal{K}_1}p_{K|B}\left(k|1\right)f_{\frac{\left(M-1\right)S^2}{\sigma^2}}\left(z;k\right).
\end{align}

The receiver observes $M$ information molecules' arrival time $Y_1,Y_2,\cdots,Y_M$, calculates their sample variance $S^2$ according to (\ref{Y_i_variance_definition_M}), and needs to give $\hat{B}$, an estimation of the information bit $B$. Given an observation $\frac{\left(M-1\right)S^2}{\sigma^2}=z$, the maximum likelihood estimation of $B$, denoted as $\hat{B}_{ML}$, is given by
\begin{align}\label{ML_receiver_design_noise}
    \hat{B}_{ML}=\arg\max_{b\in\{0,1\}}f_{\frac{\left(M-1\right)S^2}{\sigma^2}|B}\left(z|b\right).
\end{align}

\vspace{-0em}
\section{Performance Analysis of Receiver Designs}
In this section, the bit error ratio of receiver designs based on ML criterion in (\ref{ML_receiver_design}) (when background noise is negligible) and in (\ref{ML_receiver_design_noise}) (when background noise is not negligible) are both theoretically analyzed in the time-asynchronous channel model.

\vspace{-0em}
\subsection{Performance Analysis of Receiver Design when Background Noise is Negligible}
When $N$ information molecules are released at the transmitter, without being degraded by background noise in the channel environment, according to (\ref{ML_receiver_design}), the BER is given by
\begin{align}\label{BER_analysis_Pe_1}
    P_e=p_0\mathrm{Pr}\left\{0\rightarrow1\right\}+p_1\mathrm{Pr}\left\{1\rightarrow0\right\},
\end{align}
where $p_0$ and $p_1$ are the priori probability of $B=0$ and $B=1$ respectively, and $\mathrm{Pr}\left\{i\rightarrow j\right\}$ is the probability of $\hat{B}_{ML}=j$ when $B=i$. $\mathrm{Pr}\left\{0\rightarrow 1\right\}$ and $\mathrm{Pr}\left\{1\rightarrow 0\right\}$ can be derived as follows:
\begin{align}\label{BER_0_1}
    \mathrm{Pr}\left\{0\rightarrow1\right\}=\int_{\mathbb{R}_1}f_{\frac{\left(N-1\right)S^2}{\sigma^2}|B}\left(z|0\right)\mathrm{d}z,
\end{align}
\begin{align}\label{BER_1_0}
    \mathrm{Pr}\left\{1\rightarrow0\right\}=\int_{\mathbb{R}_0}f_{\frac{\left(N-1\right)S^2}{\sigma^2}|B}\left(z|1\right)\mathrm{d}z,
\end{align}
where $f_{\frac{\left(N-1\right)S^2}{\sigma^2}|B}\left(z|0\right)$ and $f_{\frac{\left(N-1\right)S^2}{\sigma^2}|B}\left(z|0\right)$ are defined in (\ref{S2_bit0_pdf}) and (\ref{S2_bit1_pdf}) respectively, $\mathbb{R}_0$ and $\mathbb{R}_1$ are the integral interval, satisfying
\begin{align}\label{integral_interval_R0}
    \mathbb{R}_0\in\Big\{z|f_{\frac{\left(N-1\right)S^2}{\sigma^2}|B}\left(z|0\right)>f_{\frac{\left(N-1\right)S^2}{\sigma^2}|B}\left(z|1\right)\Big\},
\end{align}
\begin{align}\label{integral_interval_R1}
    \mathbb{R}_1\in\Big\{z|f_{\frac{\left(N-1\right)S^2}{\sigma^2}|B}\left(z|1\right)>f_{\frac{\left(N-1\right)S^2}{\sigma^2}|B}\left(z|0\right)\Big\}.
\end{align}
When (\ref{BER_0_1}) and (\ref{BER_1_0}) are plugged into (\ref{BER_analysis_Pe_1}), the BER can be written as
\begin{align}\label{BER_analysis_Pe_2}
    P_e=p_0\int_{\mathbb{R}_1}f_{\frac{\left(N-1\right)S^2}{\sigma^2}|B}\left(z|0\right)\mathrm{d}z+p_1\int_{\mathbb{R}_0}f_{\frac{\left(N-1\right)S^2}{\sigma^2}|B}\left(z|1\right)\mathrm{d}z.
\end{align}

\vspace{-0em}
\subsection{Performance Analysis of Receiver Design when Background Noise is not Negligible}
When $N$ information molecules are released at the transmitter and each of them vanishes in the channel environment with an equal probability $P_d$ independently, the number of molecules absorbed at the receiver $M$ follows binomial distribution with the follows pmf:
\begin{align}\label{P_M_bin}
    f_M\left(m;N,P_d\right)=\binom{N}{m}{P_d}^{N-m}{\left(1-P_d\right)}^{m}.
\end{align}
Thus, the BER is given by
\begin{align}\label{BER_analysis_noise_Pe_1}
    \notag&P_e=\sum_{M=0}^{N}\binom{N}{M}{P_d}^{N-M}{\left(1-P_d\right)}^{M}\cdot\\
    \notag&\qquad\qquad\Bigg(p_0\int_{\mathbb{R}'_1}f_{\frac{\left(M-1\right)S^2}{\sigma^2}|B}\left(z|0\right)\mathrm{d}z\\
    &\qquad\qquad\qquad\quad+p_1\int_{\mathbb{R}'_0}f_{\frac{\left(M-1\right)S^2}{\sigma^2}|B}\left(z|1\right)\mathrm{d}z\Bigg),
\end{align}
where $f_{\frac{\left(M-1\right)S^2}{\sigma^2}|B}\left(z|0\right)$ and $f_{\frac{\left(M-1\right)S^2}{\sigma^2}|B}\left(z|1\right)$ are defined in (\ref{S2_k_bit0_pdf}) and (\ref{S2_k_bit1_pdf}) respectively, $\mathbb{R}'_0$ and $\mathbb{R}'_1$ are the integral interval, satisfying
\begin{align}\label{integral_interval_noise_R0}
    \mathbb{R}'_0\in\Big\{z|f_{\frac{\left(M-1\right)S^2}{\sigma^2}|B}\left(z|0\right)>f_{\frac{\left(M-1\right)S^2}{\sigma^2}|B}\left(z|1\right)\Big\},
\end{align}
\begin{align}\label{integral_interval_noise_R1}
    \mathbb{R}'_1\in\Big\{z|f_{\frac{\left(M-1\right)S^2}{\sigma^2}|B}\left(z|1\right)>f_{\frac{\left(M-1\right)S^2}{\sigma^2}|B}\left(z|0\right)\Big\}.
\end{align}

\vspace{-0em}
\section{Numerical Results}
In this section, the error probability performance of the proposed binary asynchronous receiver designs based on the sample variance of information molecules' arrival time in the time-asynchronous channel are discussed through numerical methods, compared with the previous asynchronous receiver designs in \cite{farsad2016impact,farsad2017communication}.

\begin{table}
\vspace{-0em}
\footnotesize
\setlength{\abovecaptionskip}{1em}
\setlength{\belowcaptionskip}{0em}
\begin{center}
\caption{Simulation Parameters.}\label{parameters}
\begin{tabular}{*{4}{c}}
\toprule
    Parameters & Values\\
\midrule
    Temperature (body temperature) $T_A$ & $310\mathrm{K}$\\
    Viscosity of blood at body temperature $\eta$ & $2.46\mathrm{mPa\cdot s}$\\
    Radius of information molecules (hexoses) $r$ & $0.38\mathrm{nm}$\\
    Diffusion coefficient of hexoses in blood $D$ & $242.78\mu\mathrm{m^2/s}$\\
    Distance from TX to Rx in capillaries $d_c$ & $7.9\times10^2\mu\mathrm{m}$\\
    Drift velocity in capillaries $v_c$ & $7.9\times10^2\mathrm{\mu m/s}$\\
    Distance from TX to Rx in superior vena cavae $d_s$ & $1.2\times10^5\mu\mathrm{m}$\\
    Drift velocity in superior vena cavae $v_s$ & $1.2\times10^5\mathrm{\mu m/s}$\\
\bottomrule
\end{tabular}
\end{center}
\vspace{-0em}
\end{table}

In the simulation, the channel environment is in the capillaries and superior vena cavae with a positive drift velocity, and hexoses are employed as information molecules. The simulation parameters are set according to \cite{kim2014symbol}, provided in Table \ref{parameters}. We assume that information bits $B$ have equal priori probability, that is, $p_0=p_1=\frac{1}{2}$; in this case, the receiver design based on the ML decision rule is of the minimum error probability. In different number of information molecules $N$, we choose the best values of $N_0$ (i.e, the number of molecules released at time $X=0$ for bit-0) and $N_1$ (i.e, the number of molecules released at time $X=0$ for bit-1) for simulations and the best selection of $N_0$ and $N_1$ is obtained by numerical methods, given in Appendix I.

\begin{figure}[!t]
\vspace{-0em}
\setlength{\abovecaptionskip}{-0pt}
\setlength{\belowcaptionskip}{-0pt}
    \centering
    \includegraphics[width=4.5in]{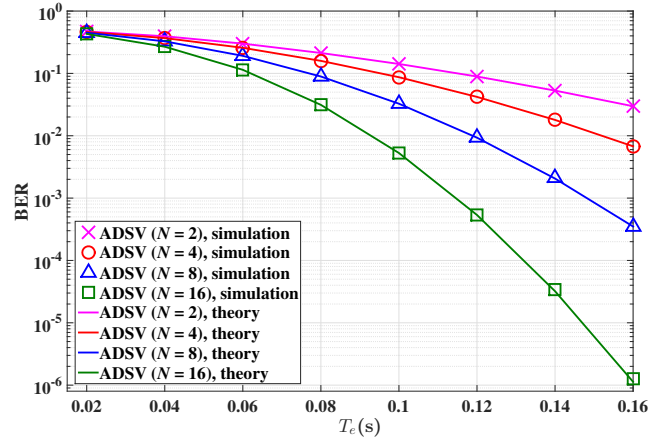}
    \caption{BER versus $T_e$ for the proposed ADSV in capillaries when background noise is negligible, where the number of molecules released per information bit is $N=2,4,8,16$ respectively.}\label{simu_fig1}
\vspace{-0em}
\end{figure}

\begin{figure}[!t]
\vspace{-0em}
\setlength{\abovecaptionskip}{-0pt}
\setlength{\belowcaptionskip}{-0pt}
    \centering
    \includegraphics[width=4.5in]{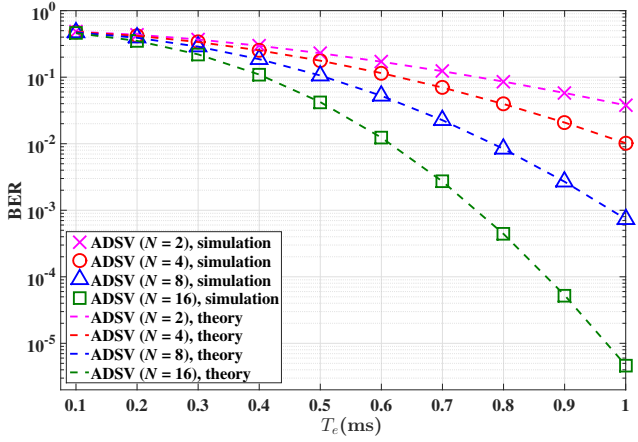}
    \caption{BER versus $T_e$ for the proposed ADSV in superior vena cavae when background noise is not negligible, where the number of molecules released per information bit is $N=2,4,8,16$ respectively.}\label{simu_fig2}
\vspace{-0em}
\end{figure}

Firstly, we investigate the case that information molecules propagate in the channel environment without degraded by background noise. The BER performance of the proposed Asynchronous Detection based on Sample Variance (ADSV) in capillaries (with short distance and low drift velocity) and super vena cavae (with long distance and high drift velocity) are shown in Fig. \ref{simu_fig1} and Fig. \ref{simu_fig2} respectively. It can be found that the numerical results agree with the theoretical analysis perfectly, and the BER performance is better with the increase of the number molecules released for each information bit, which demonstrate the feasibility of our proposed receiver designs.

\begin{figure}[!t]
\vspace{-0em}
\setlength{\abovecaptionskip}{-0pt}
\setlength{\belowcaptionskip}{-0pt}
    \centering
    \includegraphics[width=4.5in]{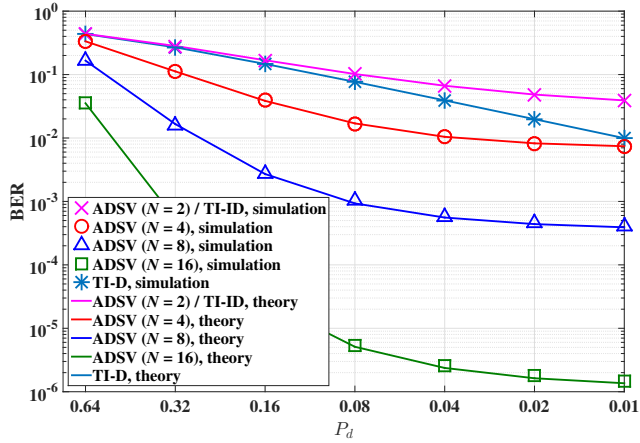}
    \caption{BER versus the degradation probability $P_d$ for the proposed ADSV and the asynchronous modulation schemes in \cite{farsad2016impact,farsad2017communication} in capillaries, where $T_e=0.16\mathrm{s}$.}\label{simu_fig3}
\vspace{-0em}
\end{figure}

\begin{figure}[!t]
\vspace{-0em}
\setlength{\abovecaptionskip}{-0pt}
\setlength{\belowcaptionskip}{-0pt}
    \centering
    \includegraphics[width=4.5in]{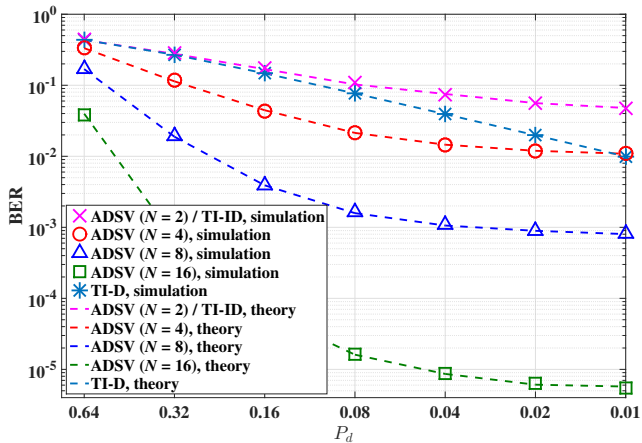}
    \caption{BER versus the degradation probability $P_d$ for the proposed ADSV and the asynchronous modulation schemes in \cite{farsad2016impact,farsad2017communication} in superior vena cavae, where $T_e=1\mathrm{ms}$.}\label{simu_fig4}
\vspace{-0em}
\end{figure}

In \cite{farsad2016impact,farsad2017communication}, an asynchronous modulation scheme was proposed, where information is modulated on the time interval between two released molecules (two indistinguishable molecules or two distinguishable molecules) and the receiver demodulates the information according to the arrival time interval of these two molecules. In this scheme, when two indistinguishable molecules are employed, it would be equivalent to our proposed asynchronous scheme with $N=2$. In Fig. \ref{simu_fig3} and Fig. \ref{simu_fig4}, the BER performance of the proposed ADSV are compared with the asynchronous modulation schemes in \cite{farsad2016impact,farsad2017communication}, including information modulated on the time interval of two indistinguishable molecules (TI-ID) and distinguishable molecules (TI-D), with information molecules degraded by background noise and each of them vanishing in the channel environment with probability $P_d$. It can be found that our proposed schemes show better noise-resistance performance with the increase of information molecules released. The reason is that in \cite{farsad2016impact,farsad2017communication}, once a molecule degrades in the channel environment, the receiver will not be able to recover information based on the time interval of molecules. However, in our schemes, when $N>2$ information molecules are released at the transmitter, although some of them would be degraded in the channel environment, the receiver can recovers information based on the sample variance of the surviving molecules' arrival time.

\begin{figure}[!t]
\vspace{-0em}
\setlength{\abovecaptionskip}{-0pt}
\setlength{\belowcaptionskip}{-0pt}
    \centering
    \includegraphics[width=4.5in]{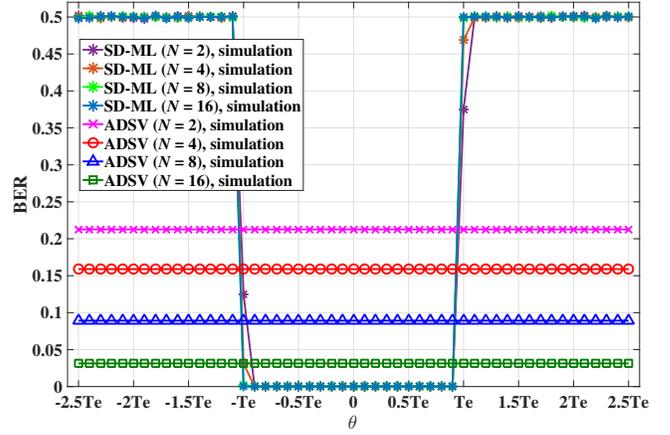}
    \caption{BER versus the clock offset $\theta$ for the synchronous ML detector (SD-ML) and the proposed ADSV, with information molecules propagation in capillaries and the background noise is negligible, where the number of molecules $N=2,4,8,16$ respectively.}\label{bias_fig1}
\vspace{-0em}
\end{figure}

Lastly, we compare the performance of the proposed asynchronous receiver designs with that of the synchronous ML detector in \cite{srinivas2012molecular}. When the clock at the receiver is perfectly synchronized with the clock at the transmitter, the synchronous ML detector can achieve very good BER performance. However, when there is an offset between these two clocks, the performance of ML detect would deteriorate. Fig. \ref{bias_fig1} shows the BER performance versus the clock offset $\theta$ for the synchronous ML detector and the proposed ADSV, with information molecules propagating in capillaries and the background being negligible. It can be found that when the clock offset $\theta$ exceeds $\pm T_e$, the performance of the proposed asynchronous receiver outperforms the synchronous ML detector. The performance of the proposed asynchronous receiver design is not effected by the clock offset because it recovers the information based on the sample variance of molecules' arrival time.

\vspace{-0em}
\section{Non-Binary Models}
In this section, we generalize the above binary models to multi-ary models. The multi-ary modulation schemes, receiver designs and the corresponding BER performance are analyzed in the time-asynchronous channel.

\vspace{-0em}
\subsection{Multi-Ary Modulation Scheme}
\begin{figure}[!t]
\vspace{-0em}
\setlength{\abovecaptionskip}{-0pt}
\setlength{\belowcaptionskip}{-0pt}
    \centering
    \includegraphics[width=2.8in]{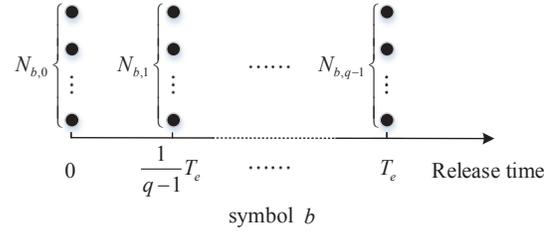}
    \caption{The diagram of the time-based $q$-ary modulation scheme for information symbol $B=b$, where $b\in\left\{0,1,\cdots,q-1\right\}$.}\label{modulation_multi_ary_fig}
\vspace{-0em}
\end{figure}

In the $q$-ary modulation scheme, information symbol $B\in\{0,1,\cdots,q-1\}$ is encoded in the release time of molecules. The proposed time-based $q$-ary modulation scheme is illustrated in Fig. \ref{modulation_multi_ary_fig}. When information symbol $b$ ($b\in\{0,1,\cdots,q-1\}$) is conveyed, $N_{b,0},N_{b,1},\cdots,N_{b,q-1}$ molecules are released at time $X=0,\frac{1}{q-1}T_e,\cdots,T_e$ respectively, where
\begin{align}
    \sum_{j=1}^{q-1}N_{b,j}=N,\ b\in\left\{0,1,\cdots,q-1\right\}.
\end{align}
The best selection of $N_{b,0},N_{b,1},\cdots,N_{b,q-1}$ (i.e., with the minimum bit error ratio) is determined by the value of $q$, $N$, $T_e$ and channel parameters, which can be derived numerically.

\vspace{-0em}
\subsection{Multi-Ary Receiver Design when Background Noise is Negligible}
\begin{theorem}\label{theorem_variance_symbol_b}
When the information conveyed is symbol $b$, denoted as $B=b$, the sampling distribution of $\frac{\left(N-1\right)S^2}{\sigma^2}$ is given by
\begin{align}\label{variance_symbol_b}
    \frac{\left(N-1\right)S^2}{\sigma^2}\thicksim\chi{'}_{N-1}^{2}\left(\lambda_b^{(N)}\right),
\end{align}
where
\begin{align}\label{sigma_bit0}
    \lambda_b^{(N)}=\frac{{T_e}^2}{N\sigma^2\left(q-1\right)^2}\sum_{j=1}^{q-1}\sum_{i=0}^{j-1}N_jN_i\left(j-i\right)^2.
\end{align}
\end{theorem}

\begin{IEEEproof}
Since the number of molecules released at time $X=0,\frac{1}{q-1}T_e,\cdots,T_e$ are $N_{b,0},N_{b,1},\cdots,N_{b,q-1}$ respectively when the information symbol is $B=b$, the random variable $X_i$ ($i\in\{1,2,\cdots,N\}$) is given by
\begin{align}
    X_i=\frac{j}{q-1}T_e,\ \mathrm{if}\ i\in\left\{\sum_{k=0}^{j}N_{b,k}-N_{b,j}+1,\cdots,\sum_{k=0}^{j}N_{b,k}\right\},
\end{align}
where $j=0,1,\cdots,q-1$. Thus, it can be deduced that
\begin{align}\label{proof_theorem_3_1}
    \notag&\sum_{i=1}^{N}\frac{\left(X_i-\overline{X}\right)^2}{\sigma^2}\\
    \notag=&\frac{1}{\sigma^2}\left(\sum_{i=1}^{N}{X_i}^2-N\overline{X}^2\right)\\
    \notag=&\frac{{T_e}^2}{\sigma^2\left(q-1\right)^2}\left(\sum_{j=0}^{q-1}N_{b,j}\cdot j^2-\frac{1}{N}\Bigg[\sum_{j=0}^{q-1}N_{b,j}\cdot j\Bigg]^2\right)\\
    \notag=&\frac{{T_e}^2}{N\sigma^2\left(q-1\right)^2}\sum_{j=1}^{q-1}\sum_{i=0}^{j-1}N_jN_i\left(j-i\right)^2\\
    =&\lambda_b^{(N)}.
\end{align}

Then according to the similar derivation in \textit{theorem} \ref{theorem_variance_bit0}, the sampling distribution of $\frac{\left(N-1\right)S^2}{\sigma^2}$ in (\ref{variance_symbol_b}) can be gotten. Thus, \textit{theorem} \ref{theorem_variance_symbol_b} is proved.
\end{IEEEproof}

The conditional PDF of $\frac{\left(N-1\right)S^2}{\sigma^2}$, given information symbol $B=b$, is
\begin{align}\label{S2_b_pdf}
    \notag& f_{\frac{\left(N-1\right)S^2}{\sigma^2}|B}\left(z|b\right)\\
    =&\frac{1}{2}e^{-{\frac{z+\lambda_b^{(N)}}{2}}}\left(\frac{z}{\lambda_b^{(N)}}\right)^{\frac{N-3}{4}}I_{\frac{N-3}{2}}\left(\sqrt{\lambda_b^{(N)}z}\right).
\end{align}

According to the sample variance of $N$ molecules' arrival time, the maximum likelihood estimation of information symbol $B$ is given by
\begin{align}\label{ML_receiver_design_multi}
    \hat{B}_{ML}=\arg\max_{b\in\{0,1,\cdots,q-1\}}f_{\frac{\left(N-1\right)S^2}{\sigma^2}|B}\left(z|b\right).
\end{align}

According to (\ref{ML_receiver_design_multi}), the symbol error probability (SEP) is given by
\begin{align}\label{BER_analysis_Pe_Multi}
    P_e=\sum_{b=0}^{q-1}p_b\Bigg(\sum_{b'\neq{b},b'=0}^{q-1}\mathrm{Pr}\left\{b\rightarrow{b'}\right\}\Bigg),
\end{align}
where $p_b$ is the priori probability of $B=b$, and $\mathrm{Pr}\left\{b\rightarrow{b'}\right\}$ is the probability of $\hat{B}_{ML}={b'}$ when $B=b$. $\mathrm{Pr}\left\{b\rightarrow{b'}\right\}$ can be derived as follows:
\begin{align}\label{SEP_b}
    \mathrm{Pr}\left\{b\rightarrow{b'}\right\}=\int_{\mathbb{R}_{b'}}f_{\frac{\left(N-1\right)S^2}{\sigma^2}|B}\left(z|b\right)\mathrm{d}z,
\end{align}
where $f_{\frac{\left(N-1\right)S^2}{\sigma^2}|B}\left(z|b\right)$ is defined in (\ref{S2_b_pdf}), $\mathbb{R}_{b'}$ is the integral interval, satisfying
\begin{align}\label{integral_interval_Rb}
    \mathbb{R}_{b'}\in\Big\{z|b'=\max_{\beta\in\left\{0,1,\cdots,q-1\right\}}f_{\frac{\left(N-1\right)S^2}{\sigma^2}|B}\left(z|\beta\right)\Big\}.
\end{align}
When (\ref{SEP_b}) is plugged into (\ref{BER_analysis_Pe_Multi}), the BER can be written as
\begin{align}\label{SEP_analysis_Pe_b}
    P_e=\sum_{b=0}^{q-1}p_b\Bigg(\sum_{b'\neq{b},b'=0}^{q-1}\int_{\mathbb{R}_{b'}}f_{\frac{\left(N-1\right)S^2}{\sigma^2}|B}\left(z|b\right)\mathrm{d}z\Bigg).
\end{align}

\vspace{-0em}
\subsection{Multi-Ary Receiver Design when Background Noise is not Negligible}
When $N$ information molecules released at the transmitter, due to the background noise is not negligible, the number of received molecules $M$ would less than $N$. Among these $M$ molecules, the numbers of molecules which released at time $X=0,\frac{1}{q-1}T_e,\cdots,T_e$ are random variables, denoted as $K_0,K_1,\cdots,K_{q-1}$ respectively and we denote their corresponding realization as:
\begin{align}
    \mathbf{k}=\left[k_0,k_1,\cdots,k_{q-1}\right].
\end{align}
The random variables $K_0,K_1,\cdots,K_{q-1}$ follow multivariate hypergeometric distribution: when the information conveyed is symbol $b$, the pmf of $K_0,K_1,\cdots,K_{q-1}$ is given by
\begin{align}\label{hypergeometric_b}
    p_{K_0,K_1,\cdots,K_{q-1}|B}\left(\mathbf{k}|b\right)=\frac{\prod_{j=0}^{q-1}\binom{N_{b,j}}{k_j}}{\binom{N}{M}},
\end{align}
where the support of $\mathbf{k}$, denoted as $\mathcal{K}_{b}$, is
\begin{align}
    \mathcal{K}_{b}:
    \left\{
        \begin{aligned}
            &k_j\in\left\{0,1,\cdots,N_{b,q-1}\right\},\ j=0,1,\cdots,q-1\\
            &\sum_{j=0}^{q-1}k_j=M.
        \end{aligned}
    \right.
\end{align}

Based on (\ref{S2_b_pdf}), it can be analogously deduced that when $M$ molecules are received, $k_j$ ($j=0,1,\cdots,q-1$) of which are released at time $X=\frac{j}{q-1}T_e$, the PDF of $\frac{\left(M-1\right)S^2}{\sigma^2}$ is
\begin{align}\label{S2_pdf_multi_M}
    \notag&f_{\frac{\left(M-1\right)S^2}{\sigma^2}}\left(z;\mathbf{k}\right)\\
    =&\frac{1}{2}e^{-{\frac{z+\lambda_{\mathbf{k}}^{(M)}}{2}}}\left(\frac{z}{\lambda_{\mathbf{k}}^{(M)}}\right)^{\frac{M-3}{4}}I_{\frac{M-3}{2}}\left(\sqrt{\lambda_{\mathbf{k}}^{(M)}z}\right),
\end{align}
where
\begin{align}
    \lambda_{\mathbf{k}}^{(M)}=\frac{{T_e}^2}{M\sigma^2\left(q-1\right)^2}\sum_{j=1}^{q-1}\sum_{i=0}^{j-1}k_jk_i\left(j-i\right)^2.
\end{align}

According to (\ref{hypergeometric_b}) and (\ref{S2_pdf_multi_M}), the conditional PDF of $\frac{\left(M-1\right)S^2}{\sigma^2}$, given $B=b$, is
\begin{align}\label{S2_symbol_pdf}
    \notag&f_{\frac{\left(M-1\right)S^2}{\sigma^2}|B}\left(z|b\right)\\ =&\sum_{\mathbf{k}\in\mathcal{K}_{b}}p_{K_0,K_1,\cdots,K_{q-1}|B}\left(\mathbf{k}|b\right)f_{\frac{\left(M-1\right)S^2}{\sigma^2}}\left(z;\mathbf{k}\right).
\end{align}

According to the sample variance of $M$ molecules¡¯ arrival time, the maximum likelihood estimation of information symbol $B$ is given by
\begin{align}\label{ML_receiver_multi_design_noise}
    \hat{B}_{ML}=\arg\max_{b\in\{0,1,\cdots,q-1\}}f_{\frac{\left(M-1\right)S^2}{\sigma^2}|B}\left(z|b\right).
\end{align}

When $N$ information molecules are released at the transmitter and each of them vanishes in the channel environment with an equal probability $P_d$ independently, the number of molecules absorbed at the receiver $M$ follows binomial distribution with the pmf in (\ref{P_M_bin}).

The SEP is given by
\begin{align}\label{SEP_analysis_noise_Pe_b}
    \notag&P_e=\sum_{M=0}^{N}\binom{N}{M}{P_d}^{N-M}{\left(1-P_d\right)}^{M}\cdot\\
    &\qquad\ \left(\sum_{b=0}^{q-1}p_b\Bigg(\sum_{b'\neq{b},b'=0}^{q-1}\int_{\mathbb{R}'_{b'}}f_{\frac{\left(M-1\right)S^2}{\sigma^2}|B}\left(z|b\right)\mathrm{d}z\Bigg)\right),
\end{align}
where $f_{\frac{\left(M-1\right)S^2}{\sigma^2}|B}\left(z|b\right)$ is defined in (\ref{S2_symbol_pdf}), $\mathbb{R}'_{b'}$ is the integral interval, satisfying
\begin{align}\label{integral_interval_noise_Rb}
    \mathbb{R'}_{b'}\in\Big\{z|b'=\max_{\beta\in\left\{0,1,\cdots,q-1\right\}}f_{\frac{\left(M-1\right)S^2}{\sigma^2}|B}\left(z|\beta\right)\Big\}.
\end{align}

\section{Conclusion}
In this work, we proposed a novel time-based modulation scheme in time-asynchronous channels for diffusion-based molecular communication systems with drift. When the drift velocity $v$, the diffusion coefficient $D$, and the distance between the transmitter and the receiver $d$ satisfy $vd\gg D$ (e.g., in blood vessels), the propagation time of information particles approaches to a normal distribution. When the background noise is negligible, the sample variance of information molecules' arrival time follows a noncentral chi-squared distribution; when the background noise is not negligible, the PDF of the sample variance of information molecules' arrival time is the linear weighting of the PDF of noncentral chi-squared distributions, where the weighting factors follows a hypergeometric distribution. According to their conditional PDF, we given the corresponding asynchronous receiver designs based on maximum likelihood criterion, without time synchronization or measuring the clock offset. The numerical results show that the BER performance of the proposed asynchronous scheme is better with the increase of the number of released information molecules. Compared with asynchronous modulation scheme proposed in \cite{farsad2016impact,farsad2017communication} where information are encoded in the time interval of released molecules, our proposed asynchronous scheme performs better when the background noise is not negligible.

The main contribution of this paper is that a modulation scheme, employing only one type of information molecules, and the corresponding detection algorithm are proposed for molecular communication systems in the time-asynchronous channel with drift when background noise is not negligible. In future work, we will explore extending the asynchronous scheme to the case of channel environment without drift.

\vspace{-0em}
\appendices

\section{The Best Value of $N_0$ and $N_1$ for the Proposed Time-Based Binary Modulation Scheme}
In Table \ref{best_N0_N1}, we utilize numerical methods to obtain the best selection of $N_0$ (i.e, the number of molecules released at time $X=0$ for bit-0) and $N_1$ (i.e, the number of molecules released at time $X=0$ for bit-1) for the proposed time-based binary modulation scheme in the time-asynchronous channel. The value of $N_0$ and $N_1$ in the above simulations are all based on this table.

\vspace{-0em}
\begin{table}
\vspace{-0em}
\footnotesize
\setlength{\abovecaptionskip}{1em}
\setlength{\belowcaptionskip}{-0em}
\begin{center}
\caption{The best value of $N_0$ and $N_1$ for the proposed time-based binary modulation scheme.}\label{best_N0_N1}
\begin{tabular}{c c c c c}
\toprule
      $N$ & $2$ & $4$ & $8$ & $16$\\
\midrule
    $N_0$ & $2(0)$ & $4(0)$ & $8(0)$ & $16(0)$\\
    $N_1$ & $1$ & $2$ & $4$ & $8$\\
\bottomrule
\end{tabular}
\end{center}
\vspace{-0em}
\end{table}

\vspace{-0em}

\bibliographystyle{IEEEtran}
\bibliography{IEEEabrv,TAMS}

\begin{thebibliography}{10}
\providecommand{\url}[1]{#1}
\csname url@samestyle\endcsname
\providecommand{\newblock}{\relax}
\providecommand{\bibinfo}[2]{#2}
\providecommand{\BIBentrySTDinterwordspacing}{\spaceskip=0pt\relax}
\providecommand{\BIBentryALTinterwordstretchfactor}{4}
\providecommand{\BIBentryALTinterwordspacing}{\spaceskip=\fontdimen2\font plus
\BIBentryALTinterwordstretchfactor\fontdimen3\font minus
  \fontdimen4\font\relax}
\providecommand{\BIBforeignlanguage}[2]{{%
\expandafter\ifx\csname l@#1\endcsname\relax
\typeout{** WARNING: IEEEtran.bst: No hyphenation pattern has been}%
\typeout{** loaded for the language `#1'. Using the pattern for}%
\typeout{** the default language instead.}%
\else
\language=\csname l@#1\endcsname
\fi
#2}}
\providecommand{\BIBdecl}{\relax}
\BIBdecl

\bibitem{zhai2018anti}
H.~Zhai, Q.~Liu, A.~V. Vasilakos, and K.~Yang, ``Anti-isi demodulation scheme
  and its experiment-based evaluation for diffusion-based molecular
  communication,'' \emph{IEEE transactions on nanobioscience}, vol.~17, no.~2,
  pp. 126--133, 2018.

\bibitem{nakano2013molecular}
T.~Nakano, A.~W. Eckford, and T.~Haraguchi, \emph{Molecular
  communication}.\hskip 1em plus 0.5em minus 0.4em\relax Cambridge University
  Press, 2013.

\bibitem{hirabayashi2011design}
M.~Hirabayashi, A.~Nishikawa, F.~Tanaka, M.~Hagiya, H.~Kojima, and K.~Oiwa,
  ``Design of molecular-based network robots-toward the environmental
  control,'' in \emph{2011 11th IEEE International Conference on
  Nanotechnology}.\hskip 1em plus 0.5em minus 0.4em\relax IEEE, 2011, pp.
  313--318.

\bibitem{nakano2012molecular}
T.~Nakano, M.~J. Moore, F.~Wei, A.~V. Vasilakos, and J.~Shuai, ``Molecular
  communication and networking: Opportunities and challenges,'' \emph{IEEE
  transactions on nanobioscience}, vol.~11, no.~2, pp. 135--148, 2012.

\bibitem{chahibi2013molecular}
Y.~Chahibi, M.~Pierobon, S.~O. Song, and I.~F. Akyildiz, ``A molecular
  communication system model for particulate drug delivery systems,''
  \emph{IEEE Transactions on biomedical engineering}, vol.~60, no.~12, pp.
  3468--3483, 2013.

\bibitem{femminella2015molecular}
M.~Femminella, G.~Reali, and A.~V. Vasilakos, ``A molecular communications
  model for drug delivery,'' \emph{IEEE transactions on nanobioscience},
  vol.~14, no.~8, pp. 935--945, 2015.

\bibitem{chude2017molecular}
U.~A. Chude-Okonkwo, R.~Malekian, B.~Maharaj, and A.~V. Vasilakos, ``Molecular
  communication and nanonetwork for targeted drug delivery: A survey,''
  \emph{IEEE Communications Surveys \& Tutorials}, vol.~19, no.~4, pp.
  3046--3096, 2017.

\bibitem{loscri2014security}
V.~Loscri, C.~Marchal, N.~Mitton, G.~Fortino, and A.~V. Vasilakos, ``Security
  and privacy in molecular communication and networking: Opportunities and
  challenges,'' \emph{IEEE transactions on nanobioscience}, vol.~13, no.~3, pp.
  198--207, 2014.

\bibitem{nakano2011molecular}
T.~Nakano and M.~Moore, ``Molecular communication paradigm overview,''
  \emph{Journal of Next Generation Information Technology}, vol.~2, no.~1,
  2011.

\bibitem{kuran2012interference}
M.~{\c{S}}. Kuran, H.~B. Yilmaz, T.~Tugcu, and I.~F. Akyildiz, ``Interference
  effects on modulation techniques in diffusion based nanonetworks,''
  \emph{Nano Communication Networks}, vol.~3, no.~1, pp. 65--73, 2012.

\bibitem{kim2013novel}
N.-R. Kim and C.-B. Chae, ``Novel modulation techniques using isomers as
  messenger molecules for nano communication networks via diffusion,''
  \emph{IEEE Journal on Selected Areas in Communications}, vol.~31, no.~12, pp.
  847--856, 2013.

\bibitem{eckford2009timing}
A.~W. Eckford, ``Timing information rates for active transport molecular
  communication,'' in \emph{International Conference on Nano-Networks}.\hskip
  1em plus 0.5em minus 0.4em\relax Springer, 2009, pp. 24--28.

\bibitem{farsad2016capacity}
N.~Farsad, Y.~Murin, A.~Eckford, and A.~Goldsmith, ``On the capacity of
  diffusion-based molecular timing channels,'' in \emph{2016 IEEE International
  Symposium on Information Theory (ISIT)}.\hskip 1em plus 0.5em minus
  0.4em\relax IEEE, 2016, pp. 1023--1027.

\bibitem{srinivas2012molecular}
K.~Srinivas, A.~W. Eckford, and R.~S. Adve, ``Molecular communication in fluid
  media: The additive inverse gaussian noise channel,'' \emph{IEEE transactions
  on information theory}, vol.~58, no.~7, pp. 4678--4692, 2012.

\bibitem{murin2017optimal}
Y.~Murin, N.~Farsad, M.~Chowdhury, and A.~Goldsmith, ``Optimal detection for
  diffusion-based molecular timing channels,'' \emph{arXiv preprint
  arXiv:1709.02977}, 2017.

\bibitem{murin2017diversity}
Y.~Murin, M.~Chowdhury, N.~Farsad, and A.~Goldsmith, ``Diversity gain of
  one-shot communication over molecular timing channels,'' in \emph{GLOBECOM
  2017-2017 IEEE Global Communications Conference}.\hskip 1em plus 0.5em minus
  0.4em\relax IEEE, 2017, pp. 1--6.

\bibitem{murin2018exploiting}
Y.~Murin, N.~Farsad, M.~Chowdhury, and A.~Goldsmith, ``Exploiting diversity in
  molecular timing channels via order statistics,'' \emph{arXiv preprint
  arXiv:1801.05567}, 2018.

\bibitem{schwab2012dynamical}
D.~J. Schwab, A.~Baetica, and P.~Mehta, ``Dynamical quorum-sensing in
  oscillators coupled through an external medium,'' \emph{Physica D: Nonlinear
  Phenomena}, vol. 241, no.~21, pp. 1782--1788, 2012.

\bibitem{taylor2009dynamical}
A.~F. Taylor, M.~R. Tinsley, F.~Wang, Z.~Huang, and K.~Showalter, ``Dynamical
  quorum sensing and synchronization in large populations of chemical
  oscillators,'' \emph{Science}, vol. 323, no. 5914, pp. 614--617, 2009.

\bibitem{mcmillen2002synchronizing}
D.~McMillen, N.~Kopell, J.~Hasty, and J.~Collins, ``Synchronizing genetic
  relaxation oscillators by intercell signaling,'' \emph{Proceedings of the
  National Academy of Sciences}, vol.~99, no.~2, pp. 679--684, 2002.

\bibitem{lin2015diffusion}
L.~Lin, C.~Yang, M.~Ma, and S.~Ma, ``Diffusion-based clock synchronization for
  molecular communication under inverse gaussian distribution,'' \emph{IEEE
  Sensors Journal}, vol.~15, no.~9, pp. 4866--4874, 2015.

\bibitem{lin2016time}
L.~Lin, J.~Zhang, M.~Ma, and H.~Yan, ``Time synchronization for molecular
  communication with drift,'' \emph{IEEE Communications Letters}, vol.~21,
  no.~3, pp. 476--479, 2016.

\bibitem{haselmayr2019impact}
W.~Haselmayr, N.~Varshney, A.~T. Asyhari, A.~Springer, and W.~Guo, ``On the
  impact of transposition errors in diffusion-based channels,'' \emph{IEEE
  Transactions on Communications}, vol.~67, no.~1, pp. 364--374, 2019.

\bibitem{lin2015asynchronous}
Y.-K. Lin, W.-A. Lin, C.-H. Lee, and P.-C. Yeh, ``Asynchronous threshold-based
  detection for quantity-type-modulated molecular communication systems,''
  \emph{IEEE Transactions on Molecular, Biological and Multi-Scale
  Communications}, vol.~1, no.~1, pp. 37--49, 2015.

\bibitem{atakan2012nanoscale}
B.~Atakan, S.~Galmes, and O.~B. Akan, ``Nanoscale communication with molecular
  arrays in nanonetworks,'' \emph{IEEE Transactions on NanoBioscience},
  vol.~11, no.~2, pp. 149--160, 2012.

\bibitem{farsad2016impact}
N.~Farsad, Y.~Murin, W.~Guo, C.-B. Chae, A.~Eckford, and A.~Goldsmith, ``On the
  impact of time-synchronization in molecular timing channels,'' in \emph{2016
  IEEE Global Communications Conference (GLOBECOM)}.\hskip 1em plus 0.5em minus
  0.4em\relax IEEE, 2016, pp. 1--6.

\bibitem{farsad2017communication}
N.~Farsad, Y.~Murin, W.~Guo, C.-B. Chae, A.~W. Eckford, and A.~Goldsmith,
  ``Communication system design and analysis for asynchronous molecular timing
  channels,'' \emph{IEEE Transactions on Molecular, Biological and Multi-Scale
  Communications}, vol.~3, no.~4, pp. 239--253, 2017.

\bibitem{garralda2011diffusion}
N.~Garralda, I.~Llatser, A.~Cabellos-Aparicio, E.~Alarc{\'o}n, and M.~Pierobon,
  ``Diffusion-based physical channel identification in molecular
  nanonetworks,'' \emph{Nano Communication Networks}, vol.~2, no.~4, pp.
  196--204, 2011.

\bibitem{you2004programmed}
L.~You, R.~S. Cox~III, R.~Weiss, and F.~H. Arnold, ``Programmed population
  control by cell--cell communication and regulated killing,'' \emph{Nature},
  vol. 428, no. 6985, p. 868, 2004.

\bibitem{chen2005artificial}
M.-T. Chen and R.~Weiss, ``Artificial cell-cell communication in yeast
  saccharomyces cerevisiae using signaling elements from arabidopsis
  thaliana,'' \emph{Nature biotechnology}, vol.~23, no.~12, p. 1551, 2005.

\bibitem{basu2005synthetic}
S.~Basu, Y.~Gerchman, C.~H. Collins, F.~H. Arnold, and R.~Weiss, ``A synthetic
  multicellular system for programmed pattern formation,'' \emph{Nature}, vol.
  434, no. 7037, p. 1130, 2005.

\bibitem{cheong2010oscillatory}
R.~Cheong and A.~Levchenko, ``Oscillatory signaling processes: the how, the why
  and the where,'' \emph{Current opinion in genetics \& development}, vol.~20,
  no.~6, pp. 665--669, 2010.

\bibitem{kim2014symbol}
N.-R. Kim, A.~W. Eckford, and C.-B. Chae, ``Symbol interval optimization for
  molecular communication with drift,'' \emph{IEEE transactions on
  nanobioscience}, vol.~13, no.~3, pp. 223--229, 2014.

\bibitem{langer2001drugs}
R.~Langer, ``Drugs on target,'' \emph{Science}, vol. 293, no. 5527, pp. 58--59,
  2001.

\bibitem{nakano2013transmission}
T.~Nakano, Y.~Okaie, and A.~V. Vasilakos, ``Transmission rate control for
  molecular communication among biological nanomachines,'' \emph{IEEE Journal
  on Selected Areas in Communications}, vol.~31, no.~12, pp. 835--846, 2013.

\bibitem{nakano2014molecular}
T.~Nakano, T.~Suda, Y.~Okaie, M.~J. Moore, and A.~V. Vasilakos, ``Molecular
  communication among biological nanomachines: A layered architecture and
  research issues,'' \emph{IEEE transactions on nanobioscience}, vol.~13,
  no.~3, pp. 169--197, 2014.

\bibitem{felicetti2014tcp}
L.~Felicetti, M.~Femminella, G.~Reali, T.~Nakano, and A.~V. Vasilakos,
  ``Tcp-like molecular communications,'' \emph{IEEE Journal on Selected Areas
  in Communications}, vol.~32, no.~12, pp. 2354--2367, 2014.

\bibitem{weiss2003genetic}
R.~Weiss, S.~Basu, S.~Hooshangi, A.~Kalmbach, D.~Karig, R.~Mehreja, and
  I.~Netravali, ``Genetic circuit building blocks for cellular computation,
  communications, and signal processing,'' \emph{Natural Computing}, vol.~2,
  no.~1, pp. 47--84, 2003.

\bibitem{mano2001logic}
M.~M. Mano and C.~R. Kime, \emph{Logic and Computer Design Fundamentals 2nd
  Edition Updated}.\hskip 1em plus 0.5em minus 0.4em\relax Chapter, 2001,
  vol.~6.

\bibitem{tavakkoli2016performance}
N.~Tavakkoli, P.~Azmi, and N.~Mokari, ``Performance evaluation and optimal
  detection of relay-assisted diffusion-based molecular communication with
  drift,'' \emph{IEEE transactions on nanobioscience}, vol.~16, no.~1, pp.
  34--42, 2016.

\bibitem{eckford2007nanoscale}
A.~W. Eckford, ``Nanoscale communication with brownian motion,'' in \emph{2007
  41st Annual Conference on Information Sciences and Systems}.\hskip 1em plus
  0.5em minus 0.4em\relax IEEE, 2007, pp. 160--165.

\bibitem{chhikara1988inverse}
R.~Chhikara, \emph{The Inverse Gaussian Distribution: Theory: Methodology, and
  Applications}.\hskip 1em plus 0.5em minus 0.4em\relax CRC Press, 1988,
  vol.~95.

\bibitem{degroot2012probability}
M.~H. DeGroot and M.~J. Schervish, \emph{Probability and statistics}.\hskip 1em
  plus 0.5em minus 0.4em\relax Pearson Education, 2012.

\bibitem{abramowitz1965handbook}
M.~Abramowitz and I.~A. Stegun, \emph{Handbook of mathematical functions: with
  formulas, graphs, and mathematical tables}.\hskip 1em plus 0.5em minus
  0.4em\relax Courier Corporation, 1965, vol.~55.

\end{thebibliography}
\end{document}